\documentclass[sigconf,nonacm]{acmart}

\usepackage{booktabs}
\usepackage{multirow}
\usepackage{array}
\usepackage{tabularx}
\usepackage{colortbl}
\usepackage[most]{tcolorbox}
\usepackage{pgfplots}
\usepackage{tikz}
\usetikzlibrary{positioning,arrows.meta}
\pgfplotsset{compat=1.18}
\usepgfplotslibrary{statistics}
\usepackage{CJKutf8}
\usepackage{float}
\usepackage{enumitem}
\setlist[itemize,enumerate]{noitemsep,leftmargin=*,topsep=0.3ex}
\newcommand{\zh}[1]{\begin{CJK}{UTF8}{gbsn}#1\end{CJK}}

\begin{document}

\title[Rhetorical Distortion Across Platforms]{Where Does Rhetorical Distortion Live? A Multi-Platform,
Multi-Language Measurement of Influencer Content}

\author{Xucheng Yu}
\affiliation{%
  \institution{University of Illinois Urbana-Champaign}
  \city{Champaign}
  \state{Illinois}
  \country{USA}}
\email{xy63@illinois.edu}

\author{Jiangping Chen}
\affiliation{%
  \institution{University of Illinois Urbana-Champaign}
  \city{Champaign}
  \state{Illinois}
  \country{USA}}
\email{jpchen@illinois.edu}

\author{Haohan Wang}
\affiliation{%
  \institution{University of Illinois Urbana-Champaign}
  \city{Champaign}
  \state{Illinois}
  \country{USA}}
\email{haohanw@illinois.edu}

\begin{abstract}
Much of the manipulation in influencer content is not false but disproportionate, framing
accurate facts as historic, urgent, or catastrophic.
Existing approaches either verify factual claims or detect clickbait and propaganda in
single-platform, single-language text; however, they overlook where such framing
originates, because they treat a post and the replies beneath it as one unit.
To address this gap, we measure rhetorical distortion with one instrument applied
identically across six platforms and two languages, classifying creator posts and their
replies separately.
Our three-tier pipeline detects five literature-grounded dimensions (Significance
Inflation, Anxiety Manufacturing, Novelty Claims, Loaded Language, and Temporal
Distortion): bilingual lexical rules propose candidates, negative filters cancel known
false positives, and an LLM verifies or relabels the candidates.
This design keeps labels transparent and comparable across platforms, calls the LLM on
only 14\% of posts, and supports validation against human annotators and a second model.
On 13,060 posts from 330 accounts on RSS/newsletters, YouTube, Bluesky, Reddit, Weibo, and
X/Twitter, platform is far more strongly associated with account-level distortion than
content category (partial $\eta^2=0.49$ vs.\ $0.09$), and X/Twitter scores highest
(DI 7.74 vs.\ 0.78 for RSS).
On X/Twitter, the top-5 replies under a post carry 3.5$\times$ the distortion of the tweet
they answer (1.5$\times$ at equal length), while creator tweets alone are
indistinguishable from YouTube and Bluesky; two human annotators and GPT-4o confirm this gap.
Measured distortion on threaded platforms thus lives largely in the audience layer, which
calls for measuring creator and reply text separately.
\end{abstract}



\begin{teaserfigure}
\centering
{\large \href{https://topcited.ai/}{TopCited.ai}}\\[10pt]
\includegraphics[width=\textwidth]{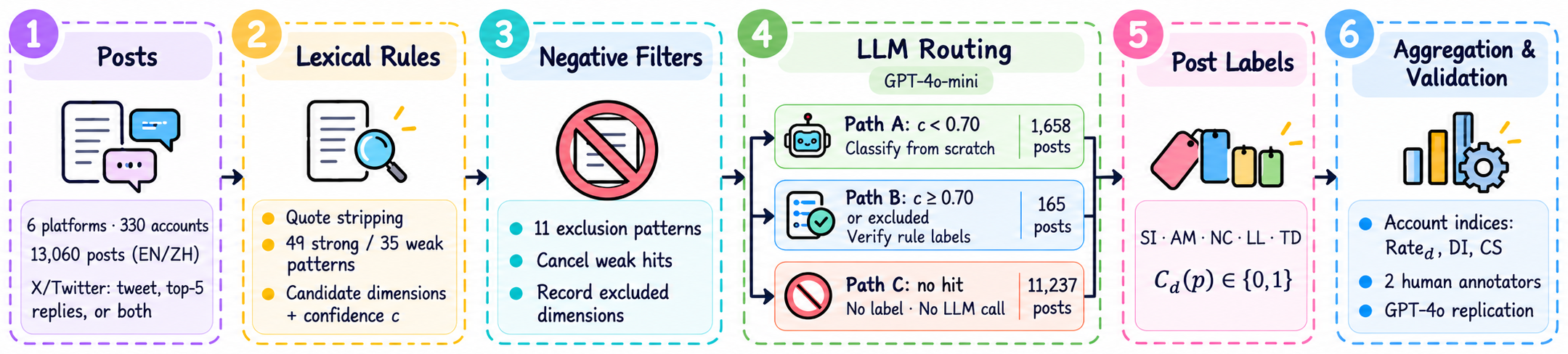}
\caption{Overview of the measurement pipeline. Lexical rules propose candidate dimensions
and a confidence $c$, negative filters cancel weak matches in known false-positive
contexts, and the post is routed to one of three paths: the LLM (GPT-4o-mini) classifies
it from scratch (Path A), verifies the rule labels (Path B), or is not called (Path C).
Numbers on the paths are post counts in the corpus. Post labels are aggregated into
per-account rates, the Distortion Index (DI), and the Consistency Score (CS). On
X/Twitter, the same pipeline is run on the merged unit, on the tweet alone, and on its
replies alone.}
\Description{A left-to-right flow diagram from posts through lexical rules and negative filters to three routing paths, post labels, and account indices.}
\label{fig:teaser}
\end{teaserfigure}

\maketitle

\section{Introduction}
\label{sec:introduction}

Social media influencers have become primary information intermediaries on the Web, and
their content is governed by engagement metrics rather than editorial standards. Locating
where their framing becomes distorted is thus a Web measurement problem. Consider a
commentator who accurately reports that ``markets fell 0.3\%'' but frames it as
``CATASTROPHIC COLLAPSE---act NOW.'' No fact is false, so the post is invisible to
fact-checking \cite{vosoughi2018spread}, and nothing is withheld, so it is not clickbait
\cite{potthast2018clickbait,chen2015misleading}. Yet the framing overstates the
evidential weight of the event and tells the audience how alarmed to be. This kind of
\emph{disproportionate framing} shapes what audiences believe matters, and it is produced
at scale on every major platform.

Existing computational work either verifies the truth of claims, detects a single
structural cue such as clickbait, or labels persuasion techniques in edited news articles
\cite{dasanmartino2019finegrained,dasanmartino2020semeval,piskorski2023semeval}. These
approaches are built for single platforms or languages, and they treat the text attached
to a post as one unit. However, they overlook where disproportionate framing originates.
On threaded platforms, what a reader sees under a creator's post also includes the
replies that the platform ranks and displays. Audience text is often uncivil
\cite{coe2014uncivil}, and audience contributions continue to reshape how a post is
evaluated after it is displayed \cite{chuai2026consensus}. It is therefore unclear
whether distortion lives with the creator, the topic, the platform, or the audience.

To address this gap, we measure rhetorical distortion with one instrument applied
identically across six platforms and two languages, and we classify creator posts and
their replies separately. We ask four research questions:
\begin{itemize}
  \item RQ1: \emph{How strongly is platform associated with distortion level, relative to
  content category?}
  \item RQ2: \emph{Do platforms elevate different kinds of distortion?}
  \item RQ3: \emph{Do the patterns transfer from English-language platforms to a
  Chinese-language platform (Weibo)?}
  \item RQ4: \emph{Does community reply content carry more distortion than the creator
  posts it responds to?}
\end{itemize}

We define rhetorical distortion through five dimensions grounded in rhetoric, framing,
and fear-appeal theory, and aggregate them into an account-level Distortion Index (DI)
and a Consistency Score (CS). A three-tier pipeline detects them
(Figure~\ref{fig:teaser}): bilingual lexical rules propose candidate dimensions, negative
filters cancel matches in known false-positive contexts, and an LLM either classifies a
candidate from scratch or verifies the rule labels. Posts with no rule hit are never sent
to the LLM, which keeps the labels transparent and comparable across platforms and limits
model calls to 14\% of posts. We apply the pipeline to 13,060 posts from 330 accounts,
classifying each X/Twitter tweet and its top-5 replies jointly and separately, and
validate the labels with two human annotators and a second LLM.

Our analysis yields four findings and one methodological contribution:
\begin{itemize}
  \item Platform is far more strongly associated with distortion than content category
  ($\eta_p^2=0.49$ vs.\ $0.09$), and the category effect is confined to X/Twitter.
  X/Twitter scores highest (DI 7.74), followed by YouTube (3.30).
  \item Platforms elevate different dimensions: X/Twitter is dominated by Loaded Language
  (LL), YouTube has the highest Significance Inflation (SI), and RSS and Weibo are SI-led
  at low levels. Distortion is concentrated in a minority of accounts, except on
  X/Twitter, where it is spread across most accounts.
  \item On X/Twitter, the reply layer carries 3.5$\times$ the distortion of the creator
  tweet it answers, mostly through LL. The gap is 1.5$\times$ at equal text length and is
  confirmed by human annotators and a second LLM, and creator tweets alone are
  indistinguishable from YouTube and Bluesky.
  \item The Weibo comparison does not support a language difference once text length,
  marker calibration, and account composition are taken into account, and distorted
  Weibo posts receive modestly more engagement than the same account's other posts.
  \item We show how the rule and LLM stages interact, that an audit-driven revision of
  the rule layer doubled agreement with human labels, and what this implies for
  distortion measurement on threaded platforms.
\end{itemize}

\section{Background and Related Work}
\label{sec:related}

\paragraph{Misinformation, clickbait, and propaganda.}
Misinformation research asks whether a claim is true and how false claims spread
\cite{vosoughi2018spread}. Clickbait detection targets headlines that withhold
information to provoke clicks \cite{potthast2018clickbait,chen2015misleading}.
Fine-grained propaganda detection labels techniques such as Loaded Language and
Exaggeration in news articles \cite{dasanmartino2019finegrained,dasanmartino2020semeval},
and shared tasks have since extended this to multilingual framing and persuasion
\cite{piskorski2023semeval}. Earlier lexical work showed that emotional and certainty
markers distinguish unreliable from reliable news \cite{rashkin2017truth}. We borrow
these techniques but apply them to influencer accounts across platforms, and we add
dimensions that propaganda taxonomies do not provide, such as temporal framing and
account-level stability.

\paragraph{Exaggeration and hyperbole.}
Computational studies of exaggeration model hyperbole as a lexical and semantic
phenomenon \cite{troiano2018exaggeration} and detect exaggerated causal claims in science
communication \cite{wright2021exaggeration}. Informal logic treats argumentative
hyperbole as a fallacy of proportion \cite{kreider2022hyperbole}. Our SI dimension adopts
this proportionality view and extends it to framing that inflates importance, urgency,
novelty, or emotional intensity.

\paragraph{LLMs as annotators.}
LLMs can match or exceed crowd workers on several text-annotation tasks
\cite{gilardi2023chatgpt}, and generative models also lower the cost of producing
persuasive content at scale \cite{goldstein2023generative}. Rather than using an LLM as
a stand-alone classifier, we use it as a verifier gated by transparent rules, report how
often it overrides them, and validate its output against human annotators and a second
model.

\paragraph{Emotion, engagement, and audience text.}
Emotional and moral language increases the diffusion of content on social networks
\cite{brady2017emotion}, and posts about political out-groups receive more engagement
than posts about in-groups \cite{rathje2021outgroup}. These findings imply that platforms
whose ranking rewards engagement also reward the kinds of wording our LL and AM
dimensions capture. Research on news comment sections finds that incivility is common in
reader comments and varies with the topic and source of the article \cite{coe2014uncivil},
and recent work on X shows that audience contributions continue to reshape how a post is
evaluated after it is displayed \cite{chuai2026consensus}. That literature studies
audience text on its own. We instead compare audience text with the creator post it
answers, using the same instrument for both.

\paragraph{Platform affordances.}
Affordances are the action possibilities that a platform's design, algorithms, and norms
make available \cite{bucher2018affordances}, and platform governance shapes what content
is produced and amplified \cite{tarleton2019facebook}. Prior work rarely measures how
affordances relate to a specific rhetorical outcome across many platforms at once.

\paragraph{Research gap.}
Existing detectors are built for single platforms, single languages, or edited news, and
they treat the text attached to a post as one unit. It therefore remains unknown whether
disproportionate framing is primarily a property of topics, of platforms, or of the
audience layer that threaded platforms attach to creator posts. Answering this requires
an instrument that is applied identically across platforms and languages, that can be
run separately on creator and reply text, and whose labels are validated against human
judgment. This is the goal of our study.

\section{Data and Methods}
\label{sec:method}

\subsection{Dataset}
\label{sec:data}

We selected accounts purposively on six platforms that differ in format and interaction
design: RSS/newsletters (long-form, no replies), YouTube (title plus transcript excerpt),
Bluesky and X/Twitter (short-form, threaded), Reddit (community-moderated), and Weibo
(Chinese short-form). Accounts span ten content categories (AI/ML, finance, media,
politics, tech, science, health, culture, education, and startup) and a range of expected
distortion levels. For each account we collected its most recent 20--50 posts in July
2026: 20 per YouTube channel, up to 30 per RSS feed, and up to 50 elsewhere. RSS,
YouTube, and Bluesky were collected through public feeds and APIs. Reddit, X/Twitter, and
Weibo were collected through browser automation on publicly visible pages. For X/Twitter
we also stored the text of the top-5 replies to each post as ranked by the platform.

Of the 13,219 collected posts, 13,060 (from 330 accounts) have retrievable text and form
the corpus. The classification unit is the post text truncated to 800 characters (Weibo
posts, which average 113 characters, are not truncated). On
X/Twitter it is, by default, the tweet concatenated with its replies; in
\S\ref{sec:decompmethod} we classify the two separately. Account-level analyses use the
\emph{analysis set} of accounts with CS$\geq$0.30 and at least 10 classified posts
($N=309$ accounts, 12,503 posts; X/Twitter $n=43$). Table~\ref{tab:data} summarizes the
corpus.

\begin{table}[t]
\centering
\footnotesize
\setlength{\tabcolsep}{2.6pt}
\caption{Dataset overview. Corpus = posts with retrievable text; analysis set = accounts
with CS$\geq$0.30 and $\geq$10 posts. On X/Twitter, 1,839 corpus posts (1,541 in the
analysis set) have stored replies.}
\label{tab:data}
\begin{tabular}{lllrr}
\toprule
\textbf{Platform} & \textbf{Collection} & \textbf{Classified unit} & \textbf{Accounts} & \textbf{Posts} \\
 & & & \textbf{(corpus/anal.)} & \textbf{(corpus/anal.)} \\
\midrule
RSS & public feeds & post text & 66 / 62 & 1,262 / 1,232 \\
YouTube & API & title + transcript & 31 / 30 & 601 / 600 \\
Bluesky & API & post & 54 / 53 & 2,680 / 2,630 \\
Reddit & browser & title + body & 57 / 56 & 2,670 / 2,669 \\
Weibo & browser & post (ZH) & 69 / 65 & 3,240 / 3,222 \\
X/Twitter & browser & tweet + top-5 replies & 53 / 43 & 2,607 / 2,150 \\
\midrule
Total & & & 330 / 309 & 13,060 / 12,503 \\
\bottomrule
\end{tabular}
\end{table}

\subsection{Theoretical Framework and Indices}
\label{sec:theory}

\paragraph{Definition.}
Following framing theory \cite{entman1993framing} and the treatment of hyperbole as a
fallacy of proportion \cite{kreider2022hyperbole}, we define \emph{rhetorical distortion}
as framing whose salience, urgency, novelty, or emotional intensity exceeds what the
underlying information warrants. Our system does not verify the underlying evidence. It
detects \emph{linguistic markers} that are characteristic of such framing, so all reported
rates are rates of distortion \emph{signals}; we return to this distinction in
\S\ref{sec:limits}. We organize the markers using the Aristotelian appeals
\cite{aristotle_rhetoric} as operationalized for computational persuasion analysis
\cite{pauli2022computational}:
\begin{itemize}
  \item \textbf{SI (logos)}: language that inflates importance beyond
    evidential support, which is salience manipulation in Entman's sense. Examples: ``changes everything,'' ``once in a generation.''
  \item \textbf{AM (pathos)}: fear appeals that target the reader's own job,
    money, skills, or future, based on the threat-severity and susceptibility
    components of the Extended Parallel Process Model \cite{witte1992putting,witte2000meta}.
    Examples: ``you will be left behind,'' ``hope you're prepared.''
  \item \textbf{NC (ethos)}: claims of exclusivity, hidden knowledge, or
    epistemic priority, a form of unwarranted appeal to authority
    \cite{flanagin2014digital}. Examples: ``nobody is talking about this.''
  \item \textbf{LL (pathos)}: emotionally charged wording or typography whose
    intensity exceeds the content \cite{dasanmartino2019finegrained,weston2018rulebook}.
    Examples: ``catastrophic,'' insults, ALL-CAPS shouting, ``!!!''.
  \item \textbf{TD (logos)}: explicit urgency labels such as ``BREAKING'' or
    ``just dropped,'' operationalizing the timeliness dimension of information
    quality \cite{wang1996beyond}.
\end{itemize}

\paragraph{Coding decisions.}
Four decisions make the dimensions operational and are shared by the rules, the LLM
prompts, and the annotation codebook. First, only the author's own wording counts:
quoted speech, headlines of linked articles reported neutrally, and self-deprecating
jokes are not distortion. Second, a post may carry several dimensions, and each is
judged independently. Third, an intense adjective on its own (``insane,''
``staggering'') is LL, and it becomes SI only when the post also claims outsized
importance; factual superlatives stated with numbers (``largest IPO on record, valued at
\$3.28T'') are neither. Fourth, AM requires that the reader be the target of the threat.
The author's own fear, reports of disasters or market drops, and general political alarm
do not count, and TD requires an explicit urgency label rather than mere recency.

\paragraph{Indices.}
The pipeline outputs, for every post, a binary label for each dimension; these labels
are aggregated per account. For account $a$ with posts $p_1,\dots,p_n$ and pipeline
output $C_d(p_i)\in\{0,1\}$ for dimension $d$, the dimension rate is
$\mathrm{Rate}_d(a)=\frac{100}{n}\sum_i C_d(p_i)$ (in percent). The five rates are the
primary outcomes. The Distortion Index is their unweighted mean,
\begin{equation}
  \mathrm{DI}(a)=\tfrac{1}{5}\left(\mathrm{SI}+\mathrm{AM}+\mathrm{NC}+\mathrm{LL}+\mathrm{TD}\right),
  \label{eq:di}
\end{equation}
which uses equal weights in the absence of evidence for differential weighting
\cite{wang1996beyond}. We analyze unrounded DI; the deployed system displays it rounded
to an integer. The Consistency Score captures week-to-week stability: posts are binned by
week, $\sigma_w$ is the standard deviation of weekly any-dimension rates, and
\begin{equation}
  \mathrm{CS}(a)=\max(0,\,1-2\sigma_w),
  \label{eq:cs}
\end{equation}
with CS$=0.5$ when all posts fall in one week. CS does not enter DI; we use it to exclude
accounts whose score would rest on one atypical week.

\subsection{Three-Tier Classification Pipeline}
\label{sec:pipeline}

The pipeline is designed around a division of labor (Figure~\ref{fig:teaser}). Lexical
rules are cheap, transparent, and identical across platforms, but they cannot judge
whether a matched phrase is used disproportionately. An LLM can make that judgment, but
applying it to every post would be costly and would make the labels depend entirely on an
opaque model. We therefore let the rules decide \emph{which} posts are examined and let
the LLM decide \emph{what} the labels are. The three tiers and the routing between them are
described below.

\paragraph{Stage 1: lexical rules (candidate generation).}
The first stage proposes candidate dimensions for each post. A single bilingual rule set
is applied to every platform. Each dimension has strong patterns, which are rarely used
in a neutral sense (e.g., ``changes everything''), and weak patterns, which are
suggestive but ambiguous (e.g., ``unprecedented''). The rule set contains 49 strong, 35
weak, and 11 exclusion patterns; SI and AM have the most strong patterns (13 each) and TD
the fewest (6). Table~\ref{tab:signals} (Appendix~\ref{app:pipeline}) lists representative patterns. Any strong or weak
match makes the post a candidate for the matched dimension, and the matches are summarized
in a rule confidence
\begin{equation}
  c(p)=\min\!\left(0.95,\; 0.35\,n_s(p)+0.15\,n_w(p)\right),
  \label{eq:conf}
\end{equation}
where $n_s$ and $n_w$ are the numbers of strong and weak matches. Two preprocessing steps
implement the coding decisions. Before matching, text inside quotation marks and block
quotes is removed, so quoted speech cannot trigger a rule. ALL-CAPS shouting is detected
case-sensitively as two or more non-acronym words in capitals, with a whitelist of common
acronyms. Chinese patterns include SI ``in history''/``historic''/``milestone'', AM
``crisis''/``collapse''/``panic'', NC ``exclusive''/``revealed''/``inside story'', LL
``shocking''/``explosive''/``scandal collapse'', and TD ``breaking''/``just now''
(original terms in Appendix~\ref{app:zh}).

\paragraph{Stage 2: negative filters (false-positive suppression).}
Lexical markers are frequently used in contexts that are not distortion, and the second
stage removes these before any LLM call. Regex exclusions cover casual time references
(``this morning I\ldots''), encouragement (``never too late to learn''), technical
severity terms (``critical vulnerability''), sports contexts, and, for Chinese,
historical retrospectives (``on this day in history''), official announcements
(``officially released''), and calming language (``view it rationally'';
Appendix~\ref{app:zh}). A filter cancels the weak matches of its dimension
but not the strong ones, so a strong marker is always passed on for verification, and
the dimensions a filter excluded are recorded so that the LLM can check them for false
negatives.

\paragraph{Stage 3: LLM verification and routing.}
The third stage assigns the final labels with GPT-4o-mini. Each post follows one of three
paths, determined by the output of Stages 1 and 2:
\begin{equation}
  \mathrm{path}(p)=
  \begin{cases}
    \text{A} & \text{if a rule fired and } c(p)<0.70,\\
    \text{B} & \text{else if a rule fired or a filter excluded}\\ & \text{a dimension},\\
    \text{C} & \text{otherwise.}
  \end{cases}
  \label{eq:route}
\end{equation}
In \emph{Path A}, the rule evidence is weak, so the LLM classifies the post from scratch
without seeing the rule output, which is intended to keep it from anchoring on ambiguous
matches. In \emph{Path B}, the rule evidence is strong ($c\geq0.70$) or a dimension was
filtered, so the LLM receives
the flagged types, the matched phrases, and the excluded types, and verifies the rule
labels and checks the excluded dimensions for false negatives in one call. In
\emph{Path C}, nothing fired, and the post is returned with no label and no LLM call. The
prompts define each type with positive and negative examples in English and Chinese,
restate the coding decisions, and instruct the model that most posts carry no distortion.
All calls use temperature 0 and JSON output (Appendix~\ref{app:prompt}).

\paragraph{How the stages interact.}
Table~\ref{tab:paths} (Appendix~\ref{app:pipeline}) reports the path distribution on the 13,060 posts. 86.0\% of posts
take Path C, and no post receives a label unless a rule or a filter fired first, so the
rules act as a recall gate. When the LLM is invoked, it is decisive. Of the 1,691
rule-flagged posts, it removed all labels on 30.6\%, changed the label set on 30.8\%, and
kept it unchanged on 38.6\%. A further 38 posts that only triggered a filter received
labels after verification. Final rates therefore reflect LLM judgments on a rule-selected
subset; \S\ref{sec:stages} analyzes this interaction per dimension.

\subsection{Creator--Reply Decomposition}
\label{sec:decompmethod}

The default X/Twitter unit merges a tweet with its top-5 replies, so its labels mix
creator and audience text. To separate the two, we run the full pipeline, with identical
rules, filters, paths, and prompts, a second time on each tweet alone and a third time on
its replies alone. Comparing the two on the same posts yields a paired creator--audience
contrast for every account. Because the reply unit concatenates up to five replies, it is
longer than a tweet and has more opportunities to match a marker. We therefore also
classify two length-controlled variants of the reply unit: the first reply alone, and the
replies truncated to the length of the tweet they answer. These variants separate
per-reply distortion from the aggregation of several replies.

\subsection{Validation}
\label{sec:audit}

\paragraph{Human annotation.}
We validate the pipeline against two human annotators who labeled the five dimensions
independently, using the codebook of \S\ref{sec:theory} and blind to all system output.
The main sample contains 198 posts stratified by platform and by whether any pipeline
version flagged the post (33 per platform). Each stratum is weighted by its size in the
corpus divided by its size in the sample, so the metrics estimate corpus-level
performance. Following the codebook rule that only the author's own wording counts, the
annotators labeled each X/Twitter item for its creator tweet and not for the attached
replies. We therefore score X/Twitter items against the pipeline's tweet-only labels
(\S\ref{sec:decompmethod}) and validate the reply layer with a second sample of 60
X/Twitter reply units (the top-5 replies of the 26 sampled tweets that had replies, plus
34 posts drawn at random from the analysis set), labeled by the same two annotators.
Agreement between the annotators is substantial: Cohen's $\kappa$ for any dimension is
0.76 in both samples; per dimension it is 0.69 (SI), 0.56 (AM), 0.78 (LL), and 1.00 for
NC and TD, which are rare (6 and 4 positives), in the main sample, and 0.76 (LL) and 0.79
(AM) for replies, where SI, NC, and TD are too rare to estimate.

\paragraph{Pipeline accuracy.}
Table~\ref{tab:audit} reports agreement with each annotator, averaged over the two. TD is
the most reliable dimension (F1 0.87). SI and LL are moderate (0.50 and 0.49). For LL,
precision is high (0.71) but recall is low (0.38): the pipeline misses intense wording
that humans flag, so LL rates are conservative. NC (0.09) and AM (0.03) are not
recovered. NC is rare and the annotators and the pipeline disagree on which posts claim
exclusivity, and AM labels largely come from the LLM's own inferences (\S\ref{sec:stages}),
which the annotators rarely shared. We therefore treat NC and AM rates as exploratory and
test every main result on a DI restricted to SI, LL, and TD (\S\ref{sec:rq1},
\S\ref{sec:decomp}). On reply units the pipeline is precise but conservative
(any-dimension precision 0.89, recall 0.60), so reply distortion is, if anything,
underestimated.

\begin{table}[t]
\centering
\footnotesize
\setlength{\tabcolsep}{3.0pt}
\caption{Agreement of the final pipeline with two human annotators (mean of the two).
Creator text: 198 posts, population-weighted, X/Twitter scored on tweets. Replies: 60
X/Twitter reply units, unweighted; -- = too few positives.}
\label{tab:audit}
\begin{tabular}{llrrrrrrr}
\toprule
 & & \textbf{Any} & \textbf{Micro} & \textbf{SI} & \textbf{AM} & \textbf{NC} & \textbf{LL} & \textbf{TD} \\
\midrule
\multirow{3}{*}{Creator} & F1 & 0.56 & 0.46 & 0.50 & 0.03 & 0.09 & 0.49 & 0.87 \\
 & Prec. & 0.62 & 0.49 & 0.41 & 0.02 & 0.10 & 0.71 & 0.77 \\
 & Rec. & 0.51 & 0.43 & 0.66 & 0.06 & 0.09 & 0.38 & 1.00 \\
\midrule
\multirow{3}{*}{Replies} & F1 & 0.71 & 0.62 & -- & 0.62 & -- & 0.66 & -- \\
 & Prec. & 0.89 & 0.78 & -- & 0.50 & -- & 0.92 & -- \\
 & Rec. & 0.60 & 0.51 & -- & 0.83 & -- & 0.51 & -- \\
\bottomrule
\end{tabular}
\end{table}

\paragraph{Audit-driven revision.}
The pipeline above is the result of a revision. A first audit exposed silent defects in an
initial version of the rule layer; after fixing them, human-judged any-dimension F1 on
identical text rose from 0.24 to 0.52. Appendix~\ref{app:validation} details the defects
and the remaining error patterns.

\subsection{Estimation Models}
\label{sec:models}

\paragraph{Platform differences (RQ1, RQ3).}
Account-level DI is zero-inflated (Table~\ref{tab:main}), so we test platform differences
with a Kruskal--Wallis test, reported with $\varepsilon^2=H/(N-1)$, and Holm-corrected
pairwise Mann--Whitney tests. To compare platform and category, we fit a two-way ANOVA of
DI on platform and harmonized category (10 categories, cells with $n\ge2$, $N=300$, 41
cells). Because the design is unbalanced with empty cells, we compute Type~II sums of
squares by nested model comparison (each main effect adjusted for the other; the
interaction tested against the additive model), report partial $\eta^2$, and repeat the
test on rank-transformed DI. We report 95\% bootstrap confidence intervals (5,000
resamples of accounts) for platform means and repeat the main test under nine
combinations of CS threshold and minimum post count.

\paragraph{Text length.}
Longer texts offer more opportunities to match a pattern. We therefore fit post-level
logistic regressions of whether any dimension is flagged,
\begin{equation}
  \mathrm{logit}\,P(Y_p=1)=\beta_0+\textstyle\sum_{k}\beta_k\,\mathbb{1}[\mathrm{plat}(p)=k]+\gamma\log(\mathrm{chars}_p),
  \label{eq:logit}
\end{equation}
with RSS as the reference platform, with and without the length term $\gamma$. The same
specification, with a reply indicator in place of the platform indicators, is used at the
unit level for the creator--reply comparison.

\paragraph{Creator vs.\ reply text (RQ4).}
For each X/Twitter account we compute DI separately on its tweets and on their replies,
restricted to the same posts, and compare the paired values with Wilcoxon signed-rank
tests. Creator tweets and replies are compared with the other platforms using the same
Holm-corrected Mann--Whitney tests as above.

\paragraph{Engagement.}
For the exploratory engagement analysis (\S\ref{sec:engagement}) we estimate
\begin{equation}
  \log(1+y_p)=\alpha_{a(p)}+\beta\,D_p+\gamma\log(\mathrm{chars}_p)+\epsilon_p,
  \label{eq:fe}
\end{equation}
where $y_p$ is the repost, like, or comment count of post $p$, $D_p$ indicates that any
dimension is flagged, and $\alpha_{a(p)}$ is an account fixed effect, so that comparisons
are between posts of the same account. Standard errors are clustered by account.

\paragraph{Model robustness.}
To test whether the results depend on the LLM, we re-classify every LLM-routed post of the
analysis set, and the X/Twitter tweet and reply units, with GPT-4o under identical rules,
prompts, and paths, and compare platform orderings, effect sizes, and label agreement.

\section{Results}
\label{sec:results}

\begin{table*}[t]
\centering
\small
\caption{Account-level results by platform ($N=309$ accounts with CS$\geq$0.30 and $\geq$10 posts).
DI is the unrounded mean of the five rates; Zero = share of accounts
whose displayed (integer) DI is 0; Chars = mean length of the classified text.
X/Twitter units are tweets with their top-5 replies (see \S\ref{sec:decomp} for the split).
Bold marks the highest value per column.}
\label{tab:main}
\begin{tabular}{lrrrrrrrrrrr}
\toprule
\textbf{Platform} & \textbf{Accts} & \textbf{Posts} & \textbf{Chars} &
\textbf{DI} & \textbf{SD} & \textbf{Zero} &
\textbf{SI\%} & \textbf{AM\%} & \textbf{NC\%} & \textbf{LL\%} & \textbf{TD\%} \\
\midrule
RSS/Newsletter & 62 & 1,232 & 255 & 0.78 & 1.60 & 59.7\% & 2.49 & 0.08 & 0.24 & 1.00 & 0.08 \\
YouTube        & 30 &   600 & 419 & 3.30 & 3.43 & 16.7\% & \textbf{5.17} & 2.33 & 0.67 & 6.83 & 1.50 \\
Bluesky        & 53 & 2,630 & 191 & 1.46 & 1.46 & 28.3\% & 0.69 & 1.00 & \textbf{0.75} & 3.81 & 1.02 \\
Reddit         & 56 & 2,669 & 201 & 0.78 & 0.74 & 48.2\% & 0.47 & 0.87 & 0.11 & 2.39 & 0.07 \\
Weibo          & 65 & 3,222 & 113 & 0.97 & 1.23 & 43.1\% & 2.19 & 0.68 & 0.43 & 0.89 & 0.68 \\
X/Twitter      & 43 & 2,150 & 521 & \textbf{7.74} & 5.30 & 9.3\% & 4.98 & \textbf{9.16} & 0.74 & \textbf{21.02} & \textbf{2.79} \\
\midrule
All & 309 & 12,503 & --- & 2.15 & 3.43 & 37.5\% & --- & --- & --- & --- & --- \\
\bottomrule
\end{tabular}
\end{table*}

\subsection{Platform and Category Effects (RQ1)}
\label{sec:rq1}

\paragraph{Platform differences.}
Distortion differs sharply by platform (Table~\ref{tab:main}; Kruskal--Wallis
$H(5)=101.5$, $p<10^{-19}$, $\varepsilon^2=0.33$). X/Twitter has the highest DI (7.74,
about 10$\times$ RSS), and 90.7\% of its accounts show some signal. YouTube is second
(3.30). Pairwise, X/Twitter exceeds every other platform (Holm $p\le0.002$), and YouTube
exceeds RSS, Reddit, and Weibo ($p<0.001$) but not Bluesky ($p=0.08$). RSS, Reddit, and
Weibo form a low tier that does not differ internally (Holm $p>0.06$).

\begin{figure}[t]
\centering
\begin{tikzpicture}
\begin{axis}[
  width=\linewidth, height=4.2cm,
  boxplot/draw direction=y,
  ytick={0,5,10,15,20}, ymin=-0.5, ymax=22,
  xtick={1,2,3,4,5,6}, xticklabels={RSS,YouTube,Bluesky,Reddit,Weibo,X/Twitter},
  x tick label style={font=\scriptsize}, y tick label style={font=\scriptsize},
  ylabel={Account DI}, ylabel style={font=\small},
  axis lines*=left, ymajorgrids=true, grid style={dashed,gray!40},
]
\addplot+[boxplot prepared={lower whisker=0, lower quartile=0, median=0, upper quartile=1.0, upper whisker=2.0}, black, fill=gray!25] coordinates {(0,6.0)(0,6.0)(0,9.0)};
\addplot+[boxplot prepared={lower whisker=0, lower quartile=1.0, median=2.0, upper quartile=5.0, upper whisker=9.0}, black, fill=orange!40] coordinates {(0,14.0)};
\addplot+[boxplot prepared={lower whisker=0, lower quartile=0.4, median=1.2, upper quartile=1.94, upper whisker=4.0}, black, fill=cyan!30] coordinates {(0,4.4)(0,6.0)(0,7.2)};
\addplot+[boxplot prepared={lower whisker=0, lower quartile=0, median=0.8, upper quartile=1.2, upper whisker=2.92}, black, fill=gray!25] coordinates {};
\addplot+[boxplot prepared={lower whisker=0, lower quartile=0, median=0.8, upper quartile=1.2, upper whisker=2.8}, black, fill=gray!25] coordinates {(0,3.6)(0,5.2)(0,6.8)};
\addplot+[boxplot prepared={lower whisker=0, lower quartile=4.8, median=6.4, upper quartile=11.8, upper whisker=20.8}, black, fill=red!40] coordinates {};
\addplot[only marks, mark=diamond*, mark size=2pt, red!70!black] coordinates {(1,0.78)(2,3.30)(3,1.46)(4,0.78)(5,0.97)(6,7.74)};
\end{axis}
\end{tikzpicture}
\caption{Account-level DI by platform ($N=309$). Boxes show quartiles, whiskers
extend to 1.5 IQR, dots are outlying accounts, and diamonds mark platform means.
X/Twitter's lower quartile (4.8) lies above the upper quartile of every other
platform except YouTube.}
\label{fig:box}
\end{figure}
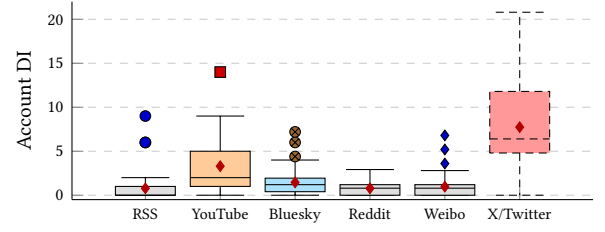

Figure~\ref{fig:box} shows that the difference is not driven by a few extreme accounts.
The median X/Twitter account (DI 6.4) scores higher than every Reddit account and all but
one account each on RSS and Weibo, whose medians are 0--0.8; the RSS and Weibo means are
pulled up by three outlying accounts each. Outside X/Twitter, YouTube has the widest
spread (IQR 1.0--5.0).

\paragraph{Platform vs.\ category.}
Content category matters far less than platform. In the two-way ANOVA, platform has a
large effect ($F(5,259)=50.16$, $p<10^{-35}$, $\eta_p^2=0.49$) and category a small one
($F(9,259)=2.96$, $p=0.002$, $\eta_p^2=0.09$), and the platform$\times$category
interaction is significant ($F(26,259)=2.05$, $p=0.003$, $\eta_p^2=0.17$). The category
effect comes from X/Twitter. Without X/Twitter, category has no effect ($F(9,224)=0.44$,
$p=0.91$) while platform remains significant ($p<10^{-8}$), and on rank-transformed DI
the category effect disappears ($p=0.41$) while platform stays large ($\eta_p^2=0.31$).
Within individual platforms, categories differ significantly only on Weibo
(Kruskal--Wallis $p=0.04$). Categories therefore differ little on most platforms but
spread widely on X/Twitter. For example, politics accounts average 12.6 on X/Twitter but
1.1--1.5 on Reddit and Bluesky, and finance accounts average 7.5 on X/Twitter but
0.8--1.1 on RSS, Reddit, and Weibo (Table~\ref{tab:pcat}, Appendix~\ref{app:rq1}). A topic, in other words,
carries very different distortion depending on where it is published.

\paragraph{Robustness.}
We next test whether the platform effect survives alternative specifications. Excluding
AM leaves the ordering unchanged (X/Twitter 7.38, YouTube 3.54, Bluesky 1.57, Weibo
1.05, RSS 0.95, Reddit 0.76; $\varepsilon^2=0.31$), and so does restricting DI to the
three dimensions with acceptable human agreement, SI, LL, and TD (\S\ref{sec:audit}):
X/Twitter 9.60, YouTube 4.50, Bluesky 1.84, Weibo 1.25, RSS 1.19, Reddit 0.98
($\varepsilon^2=0.30$). The classified text is longest on X/Twitter (521 characters) and
YouTube (419). In the post-level logistic model (Eq.~\ref{eq:logit};
Table~\ref{tab:logit}), log length is a strong predictor (OR$=2.21$). Controlling for it
shrinks the platform effects but does not remove them: relative to RSS, the odds ratio
falls from 10.3 to 6.2 for X/Twitter and from 4.1 to 2.8 for YouTube, and both remain
$p<0.001$.

\begin{table}[t]
\centering
\footnotesize
\setlength{\tabcolsep}{4pt}
\caption{Post-level logistic regression of ``any dimension flagged''
($N=13{,}060$). Odds ratios relative to RSS, without and with log text length.}
\label{tab:logit}
\begin{tabular}{lrrrr}
\toprule
 & \textbf{Flag rate} & \textbf{OR (platform)} & \textbf{OR (+length)} & $p$ \\
\midrule
RSS (ref.) & 3.4\% & 1.00 & 1.00 & -- \\
YouTube    & 12.7\% & 4.10 & 2.82 & $<$0.001 \\
Bluesky    & 6.3\% & 1.91 & 2.43 & $<$0.001 \\
Reddit     & 3.6\% & 1.05 & 1.31 & 0.16 \\
Weibo      & 4.1\% & 1.22 & 2.34 & $<$0.001 \\
X/Twitter  & 26.7\% & 10.31 & 6.20 & $<$0.001 \\
\midrule
log(chars) & & & 2.21 & $<$0.001 \\
\bottomrule
\end{tabular}
\end{table}

Bootstrap confidence intervals, nine alternative inclusion criteria ($\varepsilon^2$
between 0.29 and 0.36, same top-three order), and the rule stage alone (same top two) all
support the same ordering (Appendix~\ref{app:rq1}).

\paragraph{Model robustness.}
Re-classifying all 1,667 LLM-routed posts of the analysis set with GPT-4o (same rules,
prompts, and paths) reproduces the platform ordering exactly (X/Twitter 4.49, YouTube
2.97, Bluesky 1.06, Weibo 0.72, Reddit 0.67, RSS 0.63) with a similar effect size
($\varepsilon^2=0.28$), and account-level DI under the two models is highly correlated
(Spearman $\rho=0.90$). GPT-4o is more conservative: it flags 49.5\% of the LLM-routed
posts against 65.7\% for GPT-4o-mini, so absolute rates, unlike the ordering, depend on
the model. Label agreement between the models mirrors the human results: it is
substantial for TD and LL ($\kappa=0.70$ and 0.68), moderate for NC (0.53), and low for
SI (0.36) and AM (0.11).

\subsection{Dimension Profiles by Platform (RQ2)}
\label{sec:rq2}

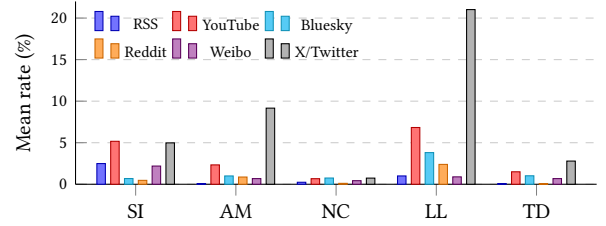
\begin{figure}[t]
\centering
\begin{tikzpicture}
\begin{axis}[
  ybar, bar width=3.2pt,
  width=\linewidth, height=4.0cm,
  symbolic x coords={SI,AM,NC,LL,TD},
  xtick=data, x tick label style={font=\small},
  ylabel={Mean rate (\%)}, ylabel style={font=\small},
  y tick label style={font=\scriptsize},
  ymin=0, ymax=22,
  axis lines*=left, ymajorgrids=true, grid style={dashed,gray!40},
  legend style={at={(0.02,0.98)},anchor=north west,font=\scriptsize,legend columns=3,draw=none,fill=none},
  enlarge x limits=0.15,
]
\addplot[fill=blue!55,draw=blue!70!black] coordinates {(SI,2.49)(AM,0.08)(NC,0.24)(LL,1.00)(TD,0.08)};
\addplot[fill=red!55,draw=red!70!black] coordinates {(SI,5.17)(AM,2.33)(NC,0.67)(LL,6.83)(TD,1.50)};
\addplot[fill=cyan!55,draw=cyan!70!black] coordinates {(SI,0.69)(AM,1.00)(NC,0.75)(LL,3.81)(TD,1.02)};
\addplot[fill=orange!65,draw=orange!80!black] coordinates {(SI,0.47)(AM,0.87)(NC,0.11)(LL,2.39)(TD,0.07)};
\addplot[fill=violet!50,draw=violet!70!black] coordinates {(SI,2.19)(AM,0.68)(NC,0.43)(LL,0.89)(TD,0.68)};
\addplot[fill=gray!60,draw=black] coordinates {(SI,4.98)(AM,9.16)(NC,0.74)(LL,21.02)(TD,2.79)};
\legend{RSS,YouTube,Bluesky,Reddit,Weibo,X/Twitter}
\end{axis}
\end{tikzpicture}
\caption{Mean dimension rates by platform ($N=309$). X/Twitter is
LL-led, and YouTube has the highest SI.}
\label{fig:dims}
\end{figure}

Having established that platforms differ in overall level, we ask \emph{which}
dimensions they elevate (Figure~\ref{fig:dims}). X/Twitter is dominated by Loaded
Language (21.0\%) and AM (9.2\%). YouTube is led by LL (6.8\%) and has the highest SI
(5.2\%), which is consistent with titles and openings that sell a video as important. RSS
shows mainly SI (2.5\%) and almost no LL (1.0\%), Bluesky and Reddit are LL-led at low
levels, and Weibo is SI-led (2.2\%); we qualify the Weibo pattern in \S\ref{sec:rq3}.

\paragraph{Co-occurrence and concentration.}
Of the 1,212 flagged posts, 29.3\% carry more than one dimension, most often AM with LL
(Appendix~\ref{app:rq2}). Distortion is concentrated in a minority of accounts (Figure~\ref{fig:lorenz}). The top
10\% of accounts (31) produce 50.5\% of all flagged posts, 28.5\% of accounts produce
none, and the Gini coefficient of account DI is 0.68. Within platforms, concentration is
highest on RSS (Gini 0.77), where a few newsletters account for nearly all signal, and
lowest on X/Twitter (0.38), where distortion is spread across most accounts. On
X/Twitter, in other words, distortion is a property of the platform's ordinary output
rather than of a few outliers.

\paragraph{Account signatures and stability.}
$k$-means on the standardized five rates separates a large low-distortion group (278
accounts) from a high-distortion tail of 31 accounts, 24 of them on X/Twitter, that
combines high LL, AM, and SI. Twelve of the 13 accounts with displayed DI $\geq$12 are on
X/Twitter, including news organizations whose own tweets are nearly neutral
(\S\ref{sec:decomp}). Most accounts have stable weekly rates (mean CS 0.73), and CS
correlates negatively with DI (Spearman $\rho=-0.44$), partly by construction: high-DI
accounts are the least predictable from week to week (details in Appendix~\ref{app:rq2}).

\subsection{Cross-Linguistic Transfer (RQ3)}
\label{sec:rq3}

Weibo sits in the low tier (DI 0.97, indistinguishable from RSS and Reddit), with an
SI-led profile. Within matched categories, Weibo SI is lower than English for finance
(1.90\% vs.\ 4.03\%; 18 vs.\ 31 accounts, excluding crypto) but higher for AI/ML (7.20\%
vs.\ 1.68\%; 5 vs.\ 26 accounts). We do not interpret either comparison as a language
difference, for three reasons. First, Weibo posts are short (113 characters), and after
controlling for length Weibo's odds of a flag exceed those of RSS (OR 2.34;
Table~\ref{tab:logit}), reversing the raw ordering. Second, the error analysis (Appendix~\ref{app:validation}) found that
Chinese superlatives (``in history'', ``historic'') are a frequent source of SI false positives
on factual news. Third, the Weibo sample is weighted toward official media and
institutional accounts. Cross-linguistic comparison therefore requires language-specific
calibration, which we leave to future work.

\subsection{Creator--Reply Decomposition on X/Twitter (RQ4)}
\label{sec:decomp}

The results so far treat each X/Twitter post as one unit that merges the tweet with its
replies. We now apply the decomposition of \S\ref{sec:decompmethod} to locate the signal.

\paragraph{Creator tweets vs.\ replies.}
On the same posts, replies score 7.65 and tweets 2.17, a 3.5$\times$ difference
(Wilcoxon signed-rank $p=2.3\times10^{-7}$; Table~\ref{tab:decomp}); replies exceed tweets
for 36 of 43 accounts and in every category (Figure~\ref{fig:replies}, Appendix~\ref{app:rq4}). The gap is driven
by Loaded Language (4.0\% vs.\ 25.0\%) and AM (1.6\% vs.\ 8.6\%); SI is modestly higher in
replies, NC does not differ, and TD is the one dimension that is higher in tweets, where
news organizations use ``BREAKING.'' Because AM is not validated by the human annotators,
we note that the LL gap alone carries the result; on a DI restricted to SI, LL, and TD,
replies score 9.68 and tweets 2.91 (Wilcoxon $p<10^{-6}$).

\begin{table}[t]
\centering
\footnotesize
\setlength{\tabcolsep}{3pt}
\caption{X/Twitter tweets vs.\ their top-5 replies, full pipeline, same posts
(43 accounts). Wilcoxon signed-rank tests.}
\label{tab:decomp}
\begin{tabular}{lrrrrrr}
\toprule
 & \textbf{DI} & \textbf{SI} & \textbf{AM} & \textbf{NC} & \textbf{LL} & \textbf{TD} \\
\midrule
Creator tweets & 2.17 & 1.81 & 1.63 & 0.48 & 3.96 & 2.95 \\
Top-5 replies  & 7.65 & 3.44 & 8.64 & 0.60 & 24.97 & 0.61 \\
\midrule
$p$ & $<$10$^{-6}$ & 0.014 & $<$10$^{-5}$ & 0.95 & $<$10$^{-6}$ & 0.058 \\
\bottomrule
\end{tabular}
\end{table}

Replies are longer than tweets (493 vs.\ 243 characters), because five replies are
concatenated. At the unit level, controlling for log length reduces the reply odds ratio
from 4.06 to 2.62, which remains highly significant ($p<10^{-17}$). The merged X/Twitter
DI (7.74) is almost entirely accounted for by the reply layer (7.65). Over all of their
tweets, including the posts without replies, creators score 2.10 (vs.\ 2.17 on the posts
that have replies). These creator tweets do not differ from YouTube (3.30) or Bluesky
(1.46) (Holm $p=0.32$), although they exceed RSS, Reddit, and Weibo ($p\le0.007$).
Replies exceed every platform ($p<10^{-4}$).

\paragraph{Human confirmation.}
Human labels confirm the gap. On the 26 posts whose tweet and replies were both labeled,
the two annotators flagged 3 and 5 tweets but 21 and 16 reply units, and no post was
flagged in its tweet but not in its replies (exact McNemar $p<10^{-5}$ and $p=0.001$).
Because the pipeline's reply labels have high precision but recall of about 0.6
(Table~\ref{tab:audit}), the measured reply DI is a conservative estimate.

\paragraph{Aggregation versus per-reply distortion.}
The reply unit concatenates up to five replies, which raises the chance that at least one
carries a marker. Table~\ref{tab:replylen} separates this aggregation effect from
per-reply distortion on the same posts. When the replies are truncated to the length of
the tweet they answer (224 vs.\ 259 characters), the reply DI falls from 7.50 to 3.24 but
still exceeds the tweet's 2.17 (1.5$\times$; higher for 29 of 43 accounts, Wilcoxon
$p=0.003$; unit-level odds ratio 1.63 controlling for log length, $p=0.004$). The excess is
concentrated in LL (12.6\% vs.\ 4.0\%). A single first reply (105 characters) has a lower
DI than the tweet (1.29, $p=0.017$) because replies almost never carry the TD and SI
framing that news accounts use in their own posts, but its LL rate is still higher (5.6\%
vs.\ 4.0\%). We therefore read the 3.5$\times$ figure as a property of the reply layer a
reader sees under a post, not of the average individual reply: per unit of text, replies
carry about 1.5 times the distortion of the tweet, almost all of it emotionally loaded
wording. With GPT-4o as classifier, the reply-layer gap is unchanged (5.08 vs.\ 1.50,
3.4$\times$; 36 of 43 accounts, $p<10^{-6}$).

\begin{table}[t]
\centering
\footnotesize
\setlength{\tabcolsep}{3pt}
\caption{Reply-length control on X/Twitter posts with replies (43 accounts,
GPT-4o-mini). Ratio and $p$ (Wilcoxon) are relative to the creator tweet.}
\label{tab:replylen}
\begin{tabular}{lrrrrr}
\toprule
\textbf{Unit} & \textbf{Chars} & \textbf{DI} & \textbf{LL\%} & \textbf{Ratio} & $p$ \\
\midrule
Creator tweet & 259 & 2.17 & 3.96 & -- & -- \\
First reply only & 105 & 1.29 & 5.61 & 0.59 & 0.017 \\
Replies, tweet length & 224 & 3.24 & 12.63 & 1.50 & 0.003 \\
Top-5 replies & 493 & 7.50 & 24.88 & 3.46 & $<10^{-6}$ \\
\bottomrule
\end{tabular}
\end{table}

\paragraph{News organizations.}
BBC News, Reuters, and the Guardian have near-zero distortion in their own tweets
(0.0--0.8) but 4.8--12.0 in the replies they attract; the New York Times is the exception,
with urgency labels in its own tweets. The same organizations score 0--1 on RSS, YouTube,
and Bluesky, which suggests that their cross-platform jump reflects the reply environment
rather than their own style (Appendix~\ref{app:rq4}).

\subsection{Distortion and Engagement (Exploratory)}
\label{sec:engagement}

Only Weibo exposes engagement counts in our data. With account fixed effects and a length
control (Eq.~\ref{eq:fe}), flagged Weibo posts receive about 20\% more reposts ($p=0.02$)
and 22\% more likes ($p=0.04$) than the same account's other posts. The effect is modest
and observational, but it is consistent with the engagement mechanism that motivates this
work \cite{brady2017emotion,rathje2021outgroup} (details in Appendix~\ref{app:engage}).

\subsection{Contribution of Pipeline Stages}
\label{sec:stages}

Because every label passes through both a rule gate and an LLM, the final rates mix two
sources of judgment (Table~\ref{tab:stages}, Appendix~\ref{app:pipeline}). Rules propose
1,623 dimension labels on 12.3\% of the posts in the analysis set; the LLM keeps 56.1\%
of them, discards the rest, and adds 521 labels the rules did not propose, so the final
flag rate is lower than that of the rules alone (8.8\% vs.\ 12.3\%). The two stages
contribute very differently by dimension. Only 7.6\% of final LL labels are LLM
additions, but 89.4\% of final AM labels are, most often on posts sent to the LLM for an
LL hit. AM, whose labels come almost entirely from unprompted LLM judgment, is also the
dimension the human annotators least often shared (F1 0.03), while TD, the most
rule-anchored dimension, agrees best with them (0.87). The AM--LL co-occurrence reported
above is therefore partly a property of the pipeline (Appendix~\ref{app:stages}).

\section{Discussion}
\label{sec:discussion}

Here, we examine where rhetorical distortion is located across six platforms and two
languages. We find that platform is strongly associated with distortion level, that
platforms elevate different dimensions, and that on X/Twitter most of the measured
distortion comes from the reply layer rather than from creators. We discuss what these
findings imply for how distortion arises, for the measurement pipeline, and for
platforms.

\paragraph{Platform-level distortion.}
The answer to RQ1 is that platform is strongly associated with distortion level and
category only weakly, with category differences confined largely to X/Twitter. The
decomposition (RQ4) refines the affordance account. On a threaded platform, most of the
measured distortion comes from the reply layer that the platform ranks and attaches to a
creator's post; X/Twitter creators themselves write at the level of YouTube and Bluesky.
Among creator-authored text, YouTube stands out, and its high SI is consistent with a
format that rewards framing a video as significant. Platform-level comparisons that merge
creator and audience text can therefore overstate creator distortion on threaded
platforms.

\paragraph{Mechanisms of the reply gap.}
Three mechanisms could produce the creator--reply gap, and our data cannot yet separate
them. First, ranking: the replies we observe are the five the platform chose to show, and
ranking that favors engagement would surface the most emotional ones
\cite{brady2017emotion}. Second, audience composition: replies to political and media
accounts draw partisan audiences, and out-group animosity is itself engaging
\cite{rathje2021outgroup}; the largest gaps in our data are indeed in politics (2.5 vs.\
12.5) and media (2.0 vs.\ 8.5). Third, norms: comment sections have long shown high
incivility independently of ranking \cite{coe2014uncivil}. The length-controlled
comparison adds a fourth, mechanical contribution: much of the reply-layer gap comes from
the aggregation of several replies, although individual replies still carry more loaded
language than the tweets they answer. Separating these mechanisms requires replies in
chronological as well as ranked order.

\paragraph{Role of the LLM.}
The stage decomposition (\S\ref{sec:stages}) suggests a practical division of labor. For
dimensions with clear lexical markers (TD, LL), the rules supply candidates and the LLM
acts as a precision filter, discarding about 40\% of them; against human labels this
combination is precise (TD 0.77, LL 0.71) though conservative for LL. For dimensions
defined by communicative intent rather than wording (AM and, to a lesser extent, SI), the
rules rarely fire, and most labels are the LLM's own inferences on posts it saw for
another reason. The human annotators rarely shared these inferences for AM. A system
aimed at AM therefore needs either intent-level training data or a dedicated LLM pass
with its own validation, rather than piggybacking on an LL gate.

\paragraph{Practical implications.}
For information literacy, the results suggest that platform-specific reading strategies
(for example, discounting reply-thread framing) may transfer better than topic-specific
skepticism. For platform governance, the reply layer is where emotionally loaded framing
concentrates around otherwise neutral news posts, which makes reply ranking a natural
point of intervention. For measurement, our experience suggests five practices for
similar systems. (1) Classify creator and audience text separately on threaded
platforms; merging them changed X/Twitter's apparent creator distortion by a factor of
3.6. Report reply-layer and per-reply measures separately, since aggregation alone more
than doubles the reply-layer rate. (2) Report dimensions individually, because the
composite hides which strategy dominates and because reliability differs sharply by
dimension. (3) Control for text length when comparing platforms, since length alone
roughly doubles the odds of a flag per unit of log length. (4) Audit the rule layer, not
only the LLM: in our case the largest errors were silent defects in the gate, invisible
from the final labels alone. (5) Validate against humans rather than another LLM, since
LLM-based audits overstated accuracy in our case.

\paragraph{Limitations.}\label{sec:limits}
The system detects linguistic markers of disproportionate framing and does not verify
that the framing exceeds the evidence. Human agreement is good for TD, moderate for SI
and LL, and poor for NC and AM, which we treat as exploratory, and the validation samples
are small. No post is labeled without a rule hit, so recall is bounded by the rule lists,
especially for Chinese. Only X/Twitter replies were collected, accounts were chosen
purposively, all findings are associational, and absolute levels depend on the LLM.
Appendix~\ref{app:discussion} discusses each limitation and future work.

\paragraph{Ethics.}
We analyze only public posts, report aggregate statistics and public account handles
without redistributing post text, and collected data at low request rates, although
some platforms' terms of service may restrict automated collection. Two research-team
members annotated public text only. Code and aggregates will be released on GitHub.

\section{Conclusion}
\label{sec:conclusion}

We introduced a five-dimension framework for rhetorical distortion and a rule-gated LLM
pipeline, and applied it to 13,060 posts from 330 accounts on six platforms; the project
is also deployed in production by \href{https://topcited.ai/}{TopCited.ai}.
Platform is far more strongly associated with distortion than topic, and on X/Twitter the
reply layer carries 3.5$\times$ the distortion of the tweet it answers, while creator
tweets resemble YouTube and Bluesky. Measuring creator and audience text separately is
therefore essential for understanding where disproportionate framing lives on threaded
platforms.

\clearpage
\bibliographystyle{ACM-Reference-Format}
\bibliography{distortion_refs}

\appendix

\section{LLM Prompts}
\label{app:prompt}

\begin{tcolorbox}[colback=gray!4!white, colframe=gray!60!black,
  title={\small\textbf{Shared type definitions (abridged)}},
  fonttitle=\small, left=3pt, right=3pt, top=3pt, bottom=3pt]
\footnotesize\ttfamily\raggedright
Label rhetorical-distortion markers in ONE social media post (English or Chinese).
Label a type ONLY when the AUTHOR's own wording clearly uses that framing device.
Neutral reporting of facts, quoted speech, and jokes about oneself are NOT distortion.
Most posts have no distortion; returning an empty list is normal.
inflate: claims something is historic or world-changing beyond the stated facts
(NO: factual records with numbers; a single intense adjective).
anxiety: tries to make the READER afraid for THEIR OWN job, money, skills, health or
future (NO: the author's own fear; reporting a disaster or market drop).
novelty: claims exclusivity, hidden knowledge, or having predicted it all along
(NO: neutral product or milestone announcements).
loaded\_language: emotionally intense words or typography beyond what the content
needs, insults, ALL-CAPS shouting, "!!!" (NO: technical severity; literal medical terms;
sports commentary; quoted speech).
temporal: explicit urgency labels such as BREAKING, "just dropped" (NO: simply
reporting recent news or dates).
\end{tcolorbox}

\noindent Path A appends: return \{"types", "confidence", "signals"\}. Path B
additionally shows the pre-filter's flagged types with matched phrases and its
excluded types, notes that keyword matches are often false positives, and returns
\{"types", "confidence", "signals", "corrections"\}. Each definition in the full
prompt includes English and Chinese positive and negative examples.

\section{Pipeline Details}
\label{app:pipeline}

This appendix gives the representative Stage-1 patterns (Table~\ref{tab:signals}), the
distribution of posts over the three routing paths (Table~\ref{tab:paths}), and the
per-dimension contribution of the rule and LLM stages (Table~\ref{tab:stages}).

\begin{table}[H]
\centering
\footnotesize
\setlength{\tabcolsep}{3pt}
\caption{Representative Stage-1 patterns (regular expressions abbreviated).}
\label{tab:signals}
\begin{tabular}{p{0.07\linewidth}p{0.43\linewidth}p{0.43\linewidth}}
\toprule
 & \textbf{Strong} & \textbf{Weak} \\
\midrule
SI & changes everything; once in a generation; never happened before; endgame
   & most important; revolutionary; unprecedented; game changer \\
AM & you'll be left behind; hope you're prepared; point of no return
   & too late; panic; fall behind; bubble \ldots\ burst \\
NC & nobody is talking about this; what the media won't tell you
   & exclusive; hidden \ldots\ truth; hard truths \\
LL & devastating; catastrophic; insane; insults; profanity
   & shocking; brutal; staggering; stupid; ALL-CAPS; \texttt{!\{2,\}} \\
TD & breaking \ldots\ news; \texttt{BREAKING}; this just in
   & just dropped; happening now; latest update \\
\bottomrule
\end{tabular}
\end{table}

\begin{table}[tbp]
\centering
\footnotesize
\setlength{\tabcolsep}{3.5pt}
\caption{Pipeline paths per platform (posts) and LLM effect on
rule-flagged posts.}
\label{tab:paths}
\begin{tabular}{lrrrrrr}
\toprule
 & \textbf{RSS} & \textbf{YT} & \textbf{BSky} & \textbf{Reddit} & \textbf{Weibo} & \textbf{X} \\
\midrule
Path C (no hit) & 1,165 & 482 & 2,384 & 2,484 & 2,953 & 1,769 \\
Path A (fresh LLM) & 92 & 106 & 288 & 171 & 210 & 791 \\
Path B (verify) & 5 & 13 & 8 & 15 & 77 & 47 \\
\midrule
Rule-flagged & 92 & 110 & 291 & 172 & 210 & 816 \\
\quad LLM cleared all & 50 & 35 & 124 & 80 & 89 & 139 \\
\quad LLM changed & 14 & 34 & 55 & 27 & 55 & 336 \\
\quad LLM kept & 28 & 41 & 112 & 65 & 66 & 341 \\
\bottomrule
\end{tabular}
\end{table}

\begin{table}[tbp]
\centering
\footnotesize
\setlength{\tabcolsep}{3.2pt}
\caption{Rule proposals versus final labels, analysis set (12,503 posts).
Top: per dimension. Bottom: share of posts with any label under the v2 pipeline, the
v3 rules alone, and the final v3 pipeline.}
\label{tab:stages}
\begin{tabular}{lrrrrr}
\toprule
 & \textbf{SI} & \textbf{AM} & \textbf{NC} & \textbf{LL} & \textbf{TD} \\
\midrule
Proposed by rules & 160 & 73 & 57 & 1,190 & 143 \\
Kept by LLM (\%) & 67.5 & 41.1 & 61.4 & 54.4 & 62.9 \\
Added by LLM & 158 & 254 & 25 & 53 & 31 \\
Final labels & 266 & 284 & 60 & 700 & 121 \\
\quad of which LLM-added (\%) & 59.4 & 89.4 & 41.7 & 7.6 & 25.6 \\
\bottomrule
\end{tabular}

\vspace{4pt}
\begin{tabular}{lrrrrrr}
\toprule
\textbf{Flag rate (\%)} & \textbf{RSS} & \textbf{YT} & \textbf{BSky} & \textbf{Reddit} & \textbf{Weibo} & \textbf{X} \\
\midrule
v2 pipeline & 2.6 & 9.7 & 3.3 & 3.0 & 1.2 & 14.0 \\
v3 rules only & 7.0 & 18.2 & 10.4 & 6.4 & 6.5 & 32.0 \\
v3 final & 3.2 & 12.5 & 6.2 & 3.6 & 4.2 & 27.4 \\
\bottomrule
\end{tabular}
\end{table}

Compared with the flag rates of the initial (v2) pipeline, the v3 rules alone flag
roughly 2--5 times as many posts on every platform, most on Weibo (1.2\% to 6.5\%) and
Bluesky (3.3\% to 10.4\%), where the v2 gate had blocked single weak signals. After LLM
verification, the final flag rate is 31--54\% below the rule-only rate outside X/Twitter
but only 14\% below it on X/Twitter. The top three platforms (X/Twitter, YouTube,
Bluesky) keep the same order under v2, under the v3 rules alone, and under the full v3
pipeline; only the order within the low tier changes. The corrections therefore changed
absolute rates and validation accuracy but not the main RQ1 result.

\subsection{Stage Contributions}
\label{app:stages}

Because every label passes through both a rule gate and an LLM, the final rates mix two
sources of judgment. Table~\ref{tab:stages} (Appendix~\ref{app:pipeline}) separates them for the 12,503 posts of the
analysis set. Rules propose 1,623 dimension labels on 12.3\% of posts. The LLM keeps
56.1\% of these proposals, discards the rest, and adds 521 labels for dimensions the
rules did not propose (435 on Path~A and 37 on Path~B posts, counted by post). The net
effect is a lower flag rate than the rules alone (8.8\% vs.\ 12.3\%) on every platform.

The two stages contribute very differently by dimension. LL is almost entirely
rule-driven: the rules propose 1,190 LL labels, the LLM keeps about half, and only 7.6\%
of final LL labels are LLM additions. AM is the opposite. The rules propose only 73 AM
labels, the LLM rejects most of them, and 89.4\% of final AM labels (254 of 284) are
dimensions the LLM added to posts that were sent to it for another reason, most often an
LL hit. On X/Twitter, 187 of 197 AM labels arise this way. SI is mixed (59.4\% added).
AM, whose labels come almost entirely from unprompted LLM judgment, is also the dimension
the human annotators least often shared (F1 0.03; \S\ref{sec:audit}), while TD, the most
rule-anchored dimension (25.6\% added), agrees best with them (0.87). This pattern also
explains the AM--LL co-occurrence reported above: because AM is usually added to posts
that already carry LL, the pairing is partly a property of the pipeline and not only of
the text. Compared with the initial (v2) pipeline, the top three platforms keep the same
order (Appendix~\ref{app:pipeline}).

\section{Validation Details}
\label{app:validation}

\paragraph{Audit-driven revision.}
The pipeline above is the result of a revision. A first audit, with an LLM annotator
(Anthropic's Claude) blind to system output, exposed defects in an initial version of the
rule layer. An ALL-CAPS pattern compiled case-insensitively matched any word of four or
more letters and fired on 78\% of posts. An ``EVERYTHING'' pattern matched any
``everything.'' A two-weak-marker requirement kept single markers such as
``unprecedented'' from ever reaching the LLM. Several generic Chinese markers
(``highest''/``largest''/``first''; Appendix~\ref{app:zh}) fired mostly on neutral news, and a filter match vetoed an entire
dimension. We fixed these defects, unified the rules across languages, and tightened the
LLM definitions. The human sample above was drawn after these changes and is disjoint
from the first audit. On its non-X/Twitter units, where both versions classify identical
text, human-judged any-dimension F1 rises from 0.24 for the initial pipeline to 0.52 for
the final one. The LLM annotator's own labels agree with the humans only moderately
(any-dimension F1 0.60--0.61 on those units), so LLM-based audits are useful for finding
defects but overstate accuracy; we report human agreement throughout.

\paragraph{Error analysis.}
The remaining errors fall into a few recurring patterns. SI false positives are mostly
factual superlatives in news (records, ``first'' awards, ``the largest model ever
released'') and Chinese markers such as ``in history'' and ``historic'' (Appendix~\ref{app:zh}) in
official reporting.
AM false negatives are personal-finance video titles that address the viewer's money
without any listed marker (``Your money is about to be worth a lot less''), which never
pass the rule gate. These patterns suggest that SI and AM need content-level rather than
lexical cues, which we discuss in \S\ref{sec:limits}.

\section{Additional Results}
\label{app:results}

\subsection{Platform and Category Effects (RQ1)}
\label{app:rq1}

\begin{table}[tbp]
\centering
\footnotesize
\setlength{\tabcolsep}{2.5pt}
\caption{Mean DI by harmonized category and platform (accounts in
parentheses; $N=309$). Chinese categories (e.g., \texttt{finance\_cn})
are merged with their English counterparts. Minor categories are merged as
follows: crypto into finance; learning into education; lifestyle, rationalism, and sports into culture;
marketing into startup; tech\_media and tech\_edu into tech.}
\label{tab:pcat}
\begin{tabular}{lrrrrrr}
\toprule
 & \textbf{RSS} & \textbf{YT} & \textbf{BSky} & \textbf{Reddit} & \textbf{Weibo} & \textbf{X} \\
\midrule
ai\_ml    & 0.2 (11) & 1.0 (3) & 0.0 (2)  & 0.7 (6)  & 2.8 (5)  & 3.6 (4) \\
culture   & 1.1 (7)  & 1.0 (1) & 2.4 (9)  & 0.6 (11) & 0.5 (12) & -- \\
education & 1.0 (2)  & 4.3 (6) & --       & 0.1 (3)  & --       & -- \\
finance   & 0.9 (12) & 5.8 (4) & 0.0 (1)  & 1.1 (12) & 0.8 (18) & 7.5 (8) \\
health    & 0.0 (3)  & --      & 1.6 (2)  & 0.0 (1)  & 0.0 (1)  & 6.4 (1) \\
media     & --       & 2.7 (3) & 1.3 (20) & 0.4 (2)  & 1.1 (11) & 7.8 (10) \\
politics  & --       & 6.0 (1) & 1.5 (6)  & 1.1 (7)  & --       & 12.6 (9) \\
science   & --       & 1.0 (2) & 1.6 (5)  & 0.9 (8)  & 0.7 (4)  & 7.4 (2) \\
startup   & 0.5 (2)  & 2.0 (1) & --       & 0.8 (1)  & 0.0 (1)  & 4.9 (3) \\
tech      & 0.9 (25) & 3.1 (9) & 1.1 (8)  & 0.6 (5)  & 1.1 (13) & 5.3 (6) \\
\bottomrule
\end{tabular}
\end{table}

Bootstrap 95\% confidence intervals agree with the pairwise tests: X/Twitter
[6.20, 9.34] is separated from all others, YouTube [2.23, 4.53] from the low tier,
Bluesky [1.10, 1.88] is intermediate, and the low tier comprises Weibo [0.70, 1.29],
Reddit [0.60, 0.99], and RSS [0.43, 1.21]. The result also does not depend on the
inclusion criteria (Table~\ref{tab:sens}). Across nine combinations of CS threshold and
minimum post count, the effect size stays between 0.29 and 0.36, the top three platforms
keep the same order, and X/Twitter's DI is 8--12 times that of RSS. Ranking the platforms
with the deterministic rule stage alone gives the same top two (X/Twitter 6.78, YouTube
4.03), so the ordering is not an artifact of the LLM.

\begin{table}[tbp]
\centering
\footnotesize
\setlength{\tabcolsep}{4pt}
\caption{Sensitivity of the platform effect to inclusion criteria. Order is by mean DI
(X = X/Twitter, Y = YouTube, B = Bluesky, W = Weibo, R = Reddit, S = RSS).}
\label{tab:sens}
\begin{tabular}{llrrlr}
\toprule
\textbf{CS} & \textbf{Min. posts} & $N$ & $\varepsilon^2$ & \textbf{Order} & \textbf{X/RSS} \\
\midrule
$\geq$0.0 & 1  & 330 & 0.36 & X>Y>B>W>S>R & 8.3 \\
$\geq$0.0 & 10 & 320 & 0.36 & X>Y>B>W>S>R & 8.4 \\
$\geq$0.0 & 20 & 297 & 0.36 & X>Y>B>W>S>R & 8.9 \\
$\geq$0.3 & 1  & 317 & 0.34 & X>Y>B>W>R>S & 10.2 \\
$\geq$0.3 & 10 & 309 & 0.33 & X>Y>B>W>R>S & 9.9 \\
$\geq$0.3 & 20 & 287 & 0.32 & X>Y>B>W>S>R & 9.2 \\
$\geq$0.5 & 1  & 304 & 0.32 & X>Y>B>W>R>S & 11.8 \\
$\geq$0.5 & 10 & 296 & 0.31 & X>Y>B>W>R>S & 11.5 \\
$\geq$0.5 & 20 & 276 & 0.29 & X>Y>B>W>S>R & 9.2 \\
\bottomrule
\end{tabular}
\end{table}

\subsection{Dimension Profiles and Accounts (RQ2)}
\label{app:rq2}

\paragraph{Co-occurrence.}
Of the 1,212 flagged posts in the full corpus, 29.3\% carry more than one dimension. The
most frequent pair is AM with LL (187 posts), followed by LL with SI (96) and LL with TD
(38), and 60.3\% of AM posts also carry LL. Fear appeals in this corpus therefore usually
arrive in emotionally loaded wording rather than on their own; \S\ref{sec:stages} shows
that this pairing is partly a property of the pipeline. On X/Twitter, multi-dimension
posts are more common among replies (29.3\%) than among creator tweets (18.7\%), and AM
with LL accounts for most of the difference (123 vs.\ 19 pairs).

\begin{figure}[tbp]
\centering
\begin{tikzpicture}
\begin{axis}[
  width=0.82\linewidth, height=3.9cm,
  xmin=0, xmax=1, ymin=0, ymax=1,
  xlabel={Cumulative share of accounts (ascending)}, ylabel={Share of flagged posts},
  xlabel style={font=\scriptsize}, ylabel style={font=\scriptsize},
  tick label style={font=\scriptsize},
  axis lines*=left, grid=major, grid style={dashed,gray!40},
  legend style={at={(0.03,0.97)},anchor=north west,font=\scriptsize,draw=none,fill=none},
]
\addplot[thick, red!70!black, mark=*, mark size=1.2pt] coordinates
 {(0,0)(0.1,0)(0.2,0)(0.3,0.005)(0.4,0.033)(0.5,0.068)(0.6,0.125)(0.7,0.200)(0.8,0.312)(0.9,0.505)(0.95,0.673)(1,1)};
\addlegendentry{Observed}
\addplot[dashed, gray] coordinates {(0,0)(1,1)};
\addlegendentry{Equality}
\end{axis}
\end{tikzpicture}
\caption{Lorenz curve of flagged posts across the 309 accounts. The bottom 30\% of
accounts contribute 0.5\% of flagged posts; the top 10\% contribute 50.5\%.}
\label{fig:lorenz}
\end{figure}
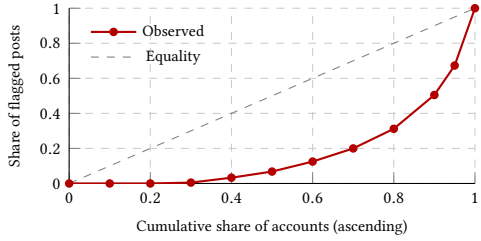

\paragraph{Account signatures.}
$k$-means on the standardized five rates separates a large low-distortion group (278
accounts) from a high-distortion tail of 31 accounts (silhouette 0.663 for $k=2$;
0.656--0.659 for $k=3,4$). The tail combines high LL (26.9\%), AM (12.1\%), and SI
(9.2\%) and consists mostly of X/Twitter accounts (24 of 31) from politics, finance, and
news media (e.g., GavinNewsom, tedcruz, jimcramer, BBCNews), plus four YouTube channels.
Table~\ref{tab:top} lists the 13 accounts with displayed DI of 12 or more. Twelve are on
X/Twitter, and for ten of them LL is the leading dimension. The list includes news
organizations whose own tweets are nearly neutral (BBCNews tweet DI 0.4, AP 2.4); for
these, the merged unit's signal comes largely from the attached replies (12.0 and 8.0;
\S\ref{sec:decomp}), which is why we caution against reading merged X/Twitter scores as
statements about authors. NASA is different: its own tweets carry TD (30\% of tweets) and
its tweet and reply DI are equal (6.0). The only non-X entry is a personal-finance
YouTube channel led by AM and TD, unlike the LL- and SI-led YouTube average.

\begin{table}[tbp]
\centering
\footnotesize
\setlength{\tabcolsep}{3pt}
\caption{Accounts with displayed DI $\geq$ 12 ($N=309$), with the leading dimension.}
\label{tab:top}
\begin{tabular}{llllr}
\toprule
\textbf{Account} & \textbf{Platform} & \textbf{Category} & \textbf{Leading dim.} & \textbf{DI} \\
\midrule
GavinNewsom   & X/Twitter & politics & LL (70\%) & 21 \\
tedcruz       & X/Twitter & politics & LL (56\%) & 18 \\
mtgreenee     & X/Twitter & politics & LL (58\%) & 18 \\
jimcramer     & X/Twitter & finance  & LL (48\%) & 16 \\
VP            & X/Twitter & politics & LL (50\%) & 16 \\
PeterSchiff   & X/Twitter & finance  & LL (36\%) & 14 \\
grahamstephan & YouTube   & finance  & AM, TD (20\%) & 14 \\
BBCNews       & X/Twitter & media    & LL (40\%) & 13 \\
CathieDWood   & X/Twitter & finance  & SI (30\%) & 12 \\
AP            & X/Twitter & media    & LL (36\%) & 12 \\
NASA          & X/Twitter & science  & TD (24\%) & 12 \\
AOC           & X/Twitter & politics & LL (38\%) & 12 \\
elonmusk      & X/Twitter & tech     & LL (36\%) & 12 \\
\bottomrule
\end{tabular}
\end{table}

\paragraph{Temporal consistency.}
Most accounts have stable weekly rates: mean CS is 0.73, and it is highest on Reddit
(0.89) and lowest on X/Twitter (0.67), excluding YouTube, where most accounts span a
single week and receive the default 0.5. CS correlates negatively with DI (Spearman
$\rho=-0.44$, $p<10^{-15}$), and the sign holds within every platform with variation
($\rho$ from $-0.29$ on Bluesky and Weibo to $-0.86$ on Reddit). Part of this is
mechanical, since CS is computed from the standard deviation of weekly any-dimension
rates, which can only be large when those rates are not near zero. It nevertheless means
that high-DI accounts are also the least predictable from week to week: their signal
varies between weeks rather than appearing as a constant style. We did not examine what
drives this variation. A single snapshot of such an account can therefore overstate or
understate its typical behavior, and the CS threshold we apply is a deliberate guard
against scoring accounts on one atypical week.

\subsection{Creator Tweets, Replies, and News Organizations (RQ4)}
\label{app:rq4}

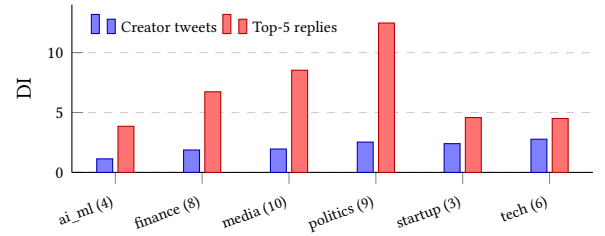
\begin{figure}[tbp]
\centering
\begin{tikzpicture}
\begin{axis}[
  ybar, bar width=6pt,
  width=\linewidth, height=3.8cm,
  symbolic x coords={ai\_ml (4),finance (8),media (10),politics (9),startup (3),tech (6)},
  xtick=data, x tick label style={font=\scriptsize,rotate=20,anchor=north east},
  ylabel={DI}, ylabel style={font=\small},
  y tick label style={font=\scriptsize},
  ymin=0, ymax=14,
  axis lines*=left, ymajorgrids=true, grid style={dashed,gray!40},
  legend style={at={(0.02,0.98)},anchor=north west,font=\scriptsize,legend columns=2,draw=none,fill=none},
  enlarge x limits=0.1,
]
\addplot[fill=blue!50,draw=blue!70!black] coordinates {(ai\_ml (4),1.13)(finance (8),1.87)(media (10),1.95)(politics (9),2.53)(startup (3),2.40)(tech (6),2.77)};
\addplot[fill=red!55,draw=red!70!black] coordinates {(ai\_ml (4),3.85)(finance (8),6.73)(media (10),8.53)(politics (9),12.47)(startup (3),4.58)(tech (6),4.50)};
\legend{Creator tweets,Top-5 replies}
\end{axis}
\end{tikzpicture}
\caption{X/Twitter creator tweets vs.\ their replies by category (full pipeline,
accounts in parentheses). Replies exceed tweets in every category.}
\label{fig:replies}
\end{figure}

\paragraph{News organizations and the same entities across platforms.}
News organizations make the pattern concrete (Table~\ref{tab:entities}). BBC News,
Reuters, and the Guardian have near-zero distortion in their own tweets (0.0--0.8) but
substantially more in the replies they attract (4.8--12.0); the same holds for AP (2.4
vs.\ 8.0), CNN (0.0 vs.\ 9.4), and the Washington Post (1.2 vs.\ 9.5). The New York Times
is the exception: its own tweets carry the signal (6.8), almost all from urgency labels
(TD on 34\% of its tweets), at the same level as its replies (6.4). Six entities in our
sample have accounts on several platforms. Every entity scores 0--1 on RSS, YouTube, and
Bluesky, whereas the four news organizations score 6--13 on X/Twitter. Two caveats apply:
the RSS feeds of news organizations are section feeds (technology) rather than the
outlet's full output, and the X/Twitter accounts of the two individuals fall below the CS
threshold. Together with the decomposition, this comparison suggests that the
cross-platform jump for organizations reflects the reply environment that X/Twitter
attaches to their posts rather than a change in their own style.

\begin{table}[tbp]
\centering
\footnotesize
\setlength{\tabcolsep}{2.8pt}
\caption{Same entity, different platform (displayed integer DI; X tweets and
replies: unrounded full-pipeline DI). $^\dagger$CS$<0.30$.}
\label{tab:entities}
\begin{tabular}{lrrrrrr}
\toprule
\textbf{Entity} & \textbf{RSS} & \textbf{YT} & \textbf{BSky} & \textbf{X} & \textbf{X tweets} & \textbf{X replies} \\
\midrule
New York Times & 0 & -- & 1 & 11 & 6.8 & 6.4 \\
BBC News & 0 & -- & -- & 13 & 0.4 & 12.0 \\
Reuters & -- & -- & 1 & 7 & 0.8 & 4.8 \\
The Guardian & 1 & -- & 0 & 6 & 0.0 & 6.0 \\
Eric Topol & -- & -- & 1 & 4$^\dagger$ & -- & -- \\
Andrej Karpathy & 0 & 0 & -- & 2$^\dagger$ & -- & -- \\
\bottomrule
\end{tabular}
\end{table}

\subsection{Distortion and Engagement (Exploratory)}
\label{app:engage}

If engagement rewards distortion, posts that carry it should attract more engagement than
other posts by the same account. Only Weibo exposes engagement counts in our data (the
X/Twitter reply count we stored is capped at the five replies collected), so this analysis
is limited to Weibo's 3,240 posts from 69 accounts, of which 134 are flagged. With account
fixed effects and a length control (Eq.~\ref{eq:fe}), flagged posts receive about 20\% more
reposts (coefficient 0.18 on log reposts, $p=0.02$) and 22\% more likes (0.20, $p=0.04$)
than the same account's other posts; the difference in comments is smaller and not
significant (0.14, $p=0.08$). The effect is modest and observational, and Weibo is a
low-distortion platform with few flagged posts. Even so, it is consistent with the
engagement mechanism that motivates this work \cite{brady2017emotion,rathje2021outgroup},
and it identifies the analysis that reply-level and engagement data on other platforms
would allow.

\section{Extended Discussion}
\label{app:discussion}

\paragraph{Limitations and future work.}
Our results point to several directions for future research.
\emph{Measurement validity.} The system detects linguistic markers of disproportionate
framing. It does not verify that framing exceeds the evidence, and a genuinely breaking
story legitimately uses ``breaking.'' Human agreement is good for TD, moderate for SI and
LL, and poor for NC and AM, so NC and AM rates are exploratory; 89\% of final AM labels
are LLM additions that no rule proposed (\S\ref{sec:stages}). The validation samples are
small (198 posts and 60 reply units, two annotators), and with 33 posts per platform they
are too small to estimate platform-level human prevalence precisely; they suggest that
pipeline recall is lower on YouTube titles than on other text.
\emph{Recall gate.} No post is labeled without a rule hit, so recall is bounded by the
rule lists; this matters most for Chinese.
\emph{Reply layer.} Only X/Twitter replies were collected, so we cannot say whether
Reddit, Bluesky, or YouTube comments show the same pattern. The top-5 replies are ranked
by the platform. The length-controlled comparison (Table~\ref{tab:replylen}) shows that
most of the reply-layer gap reflects aggregation; the per-reply excess (1.5$\times$) is
smaller and rests mainly on LL.
\emph{Sampling.} Accounts were chosen purposively, with 20--50 recent posts each, which
limits population-level inference; all findings are associational. YouTube's 20-post
windows make CS less informative there (mean 0.50).
\emph{Model dependence.} Labels depend on proprietary models whose behavior may change
across versions. The ordering and the reply gap replicate with GPT-4o, but platform-level
DI is 10--42\% lower under GPT-4o, so levels should not be compared across models.
Future work will improve AM and NC detection, collect comment layers on other platforms,
and extend Chinese marker coverage with language-specific calibration.

\section{Chinese Lexical Markers}
\label{app:zh}

Table~\ref{tab:zh} lists the original Chinese terms referred to by their English glosses
in the main text.

\begin{table}[H]
\centering
\footnotesize
\caption{Chinese terms cited in the main text, with English glosses.}
\label{tab:zh}
\setlength{\tabcolsep}{3pt}
\begin{tabularx}{\linewidth}{@{}l>{\raggedright\arraybackslash}p{2.7cm}>{\raggedright\arraybackslash}X@{}}
\toprule
Role & Chinese & Gloss \\
\midrule
SI marker & \zh{史上 / 历史性 / 里程碑} & in history / historic / milestone \\
AM marker & \zh{危机 / 崩溃 / 恐慌} & crisis / collapse / panic \\
NC marker & \zh{独家 / 揭秘 / 内幕} & exclusive / revealed / inside story \\
LL marker & \zh{震惊 / 炸裂 / 塌房} & shocking / explosive / scandal collapse \\
TD marker & \zh{突发 / 刚刚} & breaking / just now \\
\midrule
Filter: retrospective & \zh{历史上的今天} & on this day in history \\
Filter: official & \zh{官方发布} & officially released \\
Filter: calming & \zh{理性看待} & view it rationally \\
\midrule
Removed (generic) & \zh{最高 / 最大 / 首次} & highest / largest / first \\
\bottomrule
\end{tabularx}
\end{table}

\end{document}